# Time Flow and Flavor Mixing in the Flavor Spin Theory

A. Jourjine[1]

FG CTP
Dresden, Germany

**Abstract**

We show that unlike the SM in the flavor spin theories (theories with non-Euclidean/pseudo-unitary signature of the kinematic quadratic term of the Lagrangian) the Yukawa mass matrices are not arbitrary. This restricts the possible textures of both lepton and quark flavor mixing matrices to the experimentally observed form and reduces the number of real mixing parameters by two. In addition, some pairs of elements of $V_{CKM}$ and separately $U_{PMNS}$ are forced to have the same absolute values, which is experimentally confirmed within the error bounds. Apart from the flavor mixing the change in the signature imprints no changes in the SM.

[1] jourjine@pks.mpg.de

## 1. Introduction

Among the arbitrary parameters of the SM eight are the parameters of $V_{CKM}$ and $U_{PMNS}$ unitary matrices. Their textures and values remain a tantalizing flavor puzzle, despite many decades of search for an explanation. In the lepton sector $U_{PMNS}$ is close to the tribimaximal matrix. In the quark sector mixing between the first two and the third generations is by orders of magnitude smaller then between the first two [1]. In addition, a remarkable near equality of two $V_{CKM}$ elements absolute values is observed, namely $|V_{cb}| = |V_{ts}|$ within the respective 1 $\sigma$ error bounds. Assumed a coincidence, so far it received little attention.

Various methods were developed to reduce the arbitrariness of mixing parameters. A large class of models originates from and further develops the Froggart-Nielsen mechanism, where additional horizontal gauge symmetries plus a new scalar sector are introduced and then are broken to imprint a pattern on the Yukawa couplings in the form of the residual discrete groups [2]. Many continuous such as $SU(2)$, $SU(3)$ and discrete groups such as $A_4$ have been investigated in detail. More recently a modular form approach has become an alternative. For a review see [3]. Despite the plethora of the continuous and discrete groups that has been investigated there appeared no clear winner [4].

In this paper we follow a somewhat different approach applied to either three or four generations of quarks and leptons. It has some similarities with the Froggart-Nielsen mechanism. It appears naturally and is unavoidable if one tries to quantize multi-generation gauge theories whose kinematic quadratic form signature is not Euclidean but pseudo-Euclidean. In the present work the $U(3)$ invariant $(+,+,+)$ signature of the SM is replaced by either $(+,+,-)$ or $(+,+,-,-)$ signatures leading to $U(2,1)$ or $U(2,2)$ kinematic invariance of the respectively three- or four-generation theory. It turns out in such a case only a restricted set of Yukawa couplings produces real positive masses for the mass eigenstates. This results in constrains on the diagonalizing matrices and in the end on the mixing matrix parameters, eliminating one real parameter from each of $V_{CKM}$ and $U_{PMNS}$. Furthermore, all physical mass eigenstates are forced to be grouped into either doublets or singlets of the pseudo-unitary group $SU(1,1)$. Each doublet transforms under the fundamental representation of $SU(1,1)$. The singlets are split in those with the standard Dirac action and those with its negative.

We show here that the mixing matrix texture patterns for both quarks and leptons might come from the same source: the singlet/doublet assignment of the broken global $SU(2,2)$ symmetry. The corresponding local $SU(2,2)$ symmetry, together with the associated gauge field, appears as the symmetry of the kinematic part of the fermionic Lagrangian when we assume that the fermionic degrees of freedom are described by quantum differential forms. However, the theory contains no Froggart-Nielsen flavons, because the curvature of the $SU(2,2)$ connection turns out to be zero by construction. The global $SU(2,2)$ algebra is isomorphic to $SO(2,4)$, which is essentially the conformal symmetry of the massless SM. Therefore, it is broken by the same mechanisms by the mass terms and on the quantum level. As a consequence, $SU(1,1)$ of the doublets breaks down as well, because of the quantum corrections, lifting mass degeneracy of the doublets. In the end only the singlets appear on the quantum level. Some of them have the Dirac action and some have its negative.

Nevertheless the presence of the global symmetry places constraints on the physical Yukawa mass matrices. Diagonalization of the physically admissible mass matrices, together with essentially unique singlet/doublet assignments, after the symmetry breakdown results in the CKM and PMNS mixing matrices with the observed textures. Both matrices contain two real parameters and one CP violating phase, which is altogether two real parameters less then in the SM. The theory that produces the closest match with the experiment uses four generations for quarks and three or four for leptons. In addition two relations for $V_{CKM}$ and one for three-generation $U_{PMNS}$ are predicted. These are $|V_{cb}|=|V_{ts}|$, $|V_{cs}|=|V_{tb}|$ and $\sin\theta_{23}=\cos\theta_{23}$. These are satisfied within the experimental $3\sigma$ error bounds. $|V_{cb}|=|V_{ts}|$ is satisfied within the mutual $1\sigma$ errors.

Finally, we address here the issue of the quantization of the spinors with the negative action. We call them the anti-Dirac (aD) spinors. Their peculiarity is that the use of the standard quantization procedure results either in the free Hamiltonian unbounded from below or in their evolution backwards in time. We show how to cure this problem by the use of the Keldysh formalism and an appropriate assignment of the creation and annihilation operators.

The paper is organized as follows. In Section 2 we describe all possible physical eigenstates for the Yukawa mass matrices. This complements and expands the treatment in [5]. In Section 3 we describe the general setting for the transition from the interaction to the mass eigenstate basis and the resulting flavor mixing. In Section 4 we discuss in detail how the experimentally observable $V_{CKM}$ appears in our theory. In particular, we describe what happens to four-generation $V_{CKM}$ if the masses of the third and the fourth generation quarks cannot be resolved experimentally. In Section 5 we do the same for the lepton sector, describing both three- and four-generation scenarios. Section 6 is devoted to the issue of quantization of the negative action aD spinors. Section 7 is the summary.

## 2. The Physical Mass Terms

The origin of what we call the flavor spin theories lies in the bi-spinor nature of the fermionic fields extracted from inhomogeneous differential forms. The history of the attempts to use bi-spinor correspondence to differential forms is as old as the Dirac theory of the electron with the first attempt made in 1928 using a collection of antisymmetric tensors [6]. The connection between antisymmetric tensors with differential geometry and Clifford algebras was established in [7, 8]. Later the approach enjoyed a surge of interest as a basis for explaining multiple generations geometrically [9] and to describe fermions on the lattice [10]. However, the standard quantization of the theory in Minkowski space-time was shown to result in the anomalous anti-commutation relations [11], the problem that we solve in this paper, and the application of this approach was mostly in the Euclidean domain.

A different approach to the use of quantum bi-spinor fields directly, based on the extraction of the spinorial content from bi-spinors was described in detail in [12], where also the quantization of the aD doublets is also discussed. However, despite the progress in describing flavor mixing textures, the proposed solution of the problem of quantization in [12] resulted in fields propagating backwards in time. Below we will describe the correct quantization procedure. First, in this Section we will recapitulate and expand the results of [5] on the physical mass eigenstates and establish their consequences for flavor mixing.

We begin with the action for free massive fermionic bi-spinor field $\Psi \equiv \{\Psi_{\alpha\beta}\} = (\Psi_L, \Psi_R)$ given by

$$\mathcal{L} = tr\left[\overline{\overline{\Psi}}_L(i\partial)\Psi_L + \overline{\overline{\Psi}}_R(i\partial)\Psi_R - \left(\overline{\overline{\Psi}}_L \Psi_R M + \tilde{M}\,\overline{\overline{\Psi}}_R \Psi_L\right)\right], \quad (2.1)$$

where $M = M_{\alpha\beta}$ is a constant bi-spinor matrix and

$$\overline{\overline{\Psi}}_{L,R} = \gamma^0 \Psi_{L,R}^+ \gamma^0, \qquad \tilde{M} = \gamma^0 M^+ \gamma^0. \quad (2.2)$$

The usual spinors appear after a representation of bi-spinor $\Psi$ as a bilinear of two spinors. This is done by using the spinbein decomposition of $\Psi$ [5, 12]. The decomposition uses a flavor multiplet of auxiliary classical spinors. A spinbein is defined as a set of four Dirac spinors $\{\eta^A(x)\}$ that satisfy the normalization

$$\overline{\overline{\eta}}\,\eta = 1, \;\; \overline{\overline{\eta}}_\alpha^A \eta_\alpha^B = \delta^{AB}, \eta\overline{\overline{\eta}} = 1, \;\; \eta_\alpha^A \overline{\overline{\eta}}_\beta^A = \delta_{\alpha\beta}, \quad (2.3)$$

where $\overline{\overline{\eta}} \equiv \Gamma\overline{\eta}$, $\overline{\overline{\eta}}_\beta^A \equiv \Gamma^{AB}\overline{\eta}_\beta^B$, $\Gamma = diag(1,\;1,-1,-1)$. We call index $A$ interchangeably the flavor or the generation index. The spinbein decomposition is $SU(2,2)$ invariant. This is caused by the presence of $\Gamma$ in (2.3). An example of a spinbein in the momentum space is the set of familiar four normalized positive and negative energy plane wave solutions of the Dirac equation.

For a constant spinbein $\{\eta^A\}$ and a multiplet of four Dirac spinors $\psi^A(x)$ the spinbein decomposition of a bi-spinor $\Psi$ is defined by

$$\Psi = \psi\,\overline{\overline{\eta}}, \;\; \Psi_{\alpha\beta} = \psi_\alpha^A \overline{\overline{\eta}}_\beta^A, \quad \overline{\overline{\Psi}} = \eta\,\overline{\overline{\psi}}, \quad \overline{\overline{\Psi}}_{\alpha\beta} = \eta_\alpha^A \overline{\overline{\psi}}_\beta^A. \quad (2.4)$$

Using two different spinbein decompositions (2.4) of two chiral bi-spinors $\Psi_{L,R}(x)$ we obtain the reduced Lagrangian in terms of Dirac field content

$$\mathcal{L} = \overline{\psi}_L \Gamma(i\partial) \psi_L + \overline{\psi}_R \Gamma(i\partial) \psi_R - \left(\overline{\psi}_L \Gamma \mathcal{M} \psi_R + \overline{\psi}_R \tilde{\mathcal{M}} \Gamma \psi_R\right), \tag{2.5}$$

where $\mathcal{M} = \left(\overline{\overline{\eta}}_R M \eta_L\right)^T$, $\tilde{\mathcal{M}} = \Gamma \mathcal{M}^+ \Gamma$. The $SU(2,2)$ invariance of the spinbein decomposition is inherited by the kinetic term of the Lagrangian (2.5). The kinetic terms of the two of the flavors enter (2.5) with a plus, while the remaining two with a minus. The appearance of this symmetry is not surprising, considering that $SU(2,2)$ algebra is isomorphic to the algebra of the conformal group, which is the symmetry of the free massless Dirac action. In transition from bi-spinors to flavor spinors the conformal group acting on one of the indices of a bi-spinor was replaced by the $SU(2,2)$ acting on the flavor indices.

Note that the choice of normalization in (2.3) is somewhat arbitrary. If we replace in (2.3) $\eta^A$ with its unitary transform $(U\eta)^A$ in the flavor space, then instead of $\overline{\eta}\eta = \Gamma$ we obtain $U\Gamma U^+$. This can be used to obtain different forms of the Lagrangian, for example after symmetry breaking.

To extract the spinorial degrees of freedom one can also use non-constant spinbeins. Non-constant spinbeins lead to the appearance in (2.5) of a gauge field for non-compact $SU(2,2)$ gauge group. Gauge theories with non-compact gauge group fields are usually avoided in model building, because generally they lead to a non-unitary S-matrix. However, in our case, because of the normalization (2.3) the $SU(2,2)$ field has zero curvature. Therefore, it cannot carry energy and is non-dynamical. As a result, the potential $SU(2,2)$ flavons cannot appear. This can be seen as follows. In addition to making spinbein depending on the coordinate we relax the spinbein normalization in (2.3) to

$$\eta(x)\overline{\overline{\eta}}(x) = g, \quad \overline{\overline{\eta}}(x)\eta(x) = g^{-1}, \tag{2.6}$$

where $g$ is a dimensionless constant, and rescale the spinor fields $\psi \to g^{-1/2}\psi$. Now consider the kinematic term of the Lagrangian: $tr\overline{\overline{\Psi}}(i\partial)\Psi$. The substitution of (2.6) together with field redefinition transforms it into

$$tr\overline{\overline{\Psi}}(i\partial)\Psi = \overline{\overline{\psi}}(i\partial + gB_\mu)\psi, \tag{2.7}$$

where $B_\mu = \eta \partial_\mu \overline{\overline{\eta}}$, $B_\mu^{AB} = \eta_\alpha^A \partial_\mu \overline{\overline{\eta}}_\alpha^B$, is a $SU(2,2)$ gauge field in the flavor space with the coupling constant $g$. It is easy to verify that as a consequence of (2.6) its curvature is identically zero.

Returning to the original normalization (2.3), we obtain the equations of motion

$$(i\partial)\psi_L - \mathcal{M}\psi_R = 0,$$
$$(i\partial)\psi_R - \tilde{\mathcal{M}}\psi_L = 0. \tag{2.8}$$

Since in general $\tilde{\mathcal{M}} \neq \mathcal{M}^+$, it is not guaranteed that (2.8) results in physical states with real positive masses. To determine the admissible mass eigenstates, we rewrite (2.8) in the second order form to obtain separate equations on the chiral components

$$(i\partial)^2 \psi_L - \mathcal{M}\tilde{\mathcal{M}}\psi_L = 0, \qquad (i\partial)^2 \psi_R - \tilde{\mathcal{M}}\mathcal{M}\psi_R = 0. \tag{2.9}$$

For the plane wave solutions of positive and negative energy, $\psi_{L,R}(x) = \psi^0_{L,R} e^{\mp ikx}$, from (2.9) we obtain the dispersion relations for the left and the right modes

$$\det(k^2 - \mathcal{M}\tilde{\mathcal{M}}) = 0, \qquad \det(k^2 - \tilde{\mathcal{M}}\mathcal{M}) = 0. \tag{2.10}$$

The eigenvalues of $\mathcal{M}\tilde{\mathcal{M}}$, $\tilde{\mathcal{M}}\mathcal{M}$ determine the squared mass spectrum of the particles. Unlike in the SM, where $\tilde{\mathcal{M}} = \mathcal{M}^+$, matrices $\mathcal{M}\tilde{\mathcal{M}}$, $\tilde{\mathcal{M}}\mathcal{M}$ are not necessarily Hermitean and thus may have complex eigenvalues. Therefore, to deal with physical particle states, we need to determine the set of $\mathcal{M}$, $\tilde{\mathcal{M}}$ that results in the real positive particle masses.

To generate physical masses, both $\mathcal{M}\tilde{\mathcal{M}}$ and $\tilde{\mathcal{M}}\mathcal{M}$ must be hermitean and non-negative-definite. Additionally, the masses for the left and the right modes must be equal. This results in the three conditions that $\mathcal{M}, \tilde{\mathcal{M}}$ must satisfy

$$(\mathcal{M}\tilde{\mathcal{M}})^+ = \mathcal{M}\tilde{\mathcal{M}},$$
$$(\tilde{\mathcal{M}}\mathcal{M})^+ = \tilde{\mathcal{M}}\mathcal{M}, \tag{2.11}$$

and for some unitary $U$

$$\tilde{\mathcal{M}}\mathcal{M} = U^+ \mathcal{M}\tilde{\mathcal{M}} U. \tag{2.12}$$

We now solve (2.11, 2.12) for various number of generations. We begin with the two generations, where mass matrices are two dimensional. For the two dimensional case the metric in the kinematic term is $\Gamma_2 = diag(1,-1)$ instead of $\Gamma$. For $\mathcal{M}, \tilde{\mathcal{M}}$ we obtain

$$\mathcal{M} = \begin{bmatrix} a_{11} & a_{12} \\ a_{21} & a_{22} \end{bmatrix}, \qquad \tilde{\mathcal{M}} = \begin{bmatrix} a_{11}^* & -a_{21}^* \\ -a_{12}^* & a_{22}^* \end{bmatrix}, \tag{2.13}$$

$$\mathcal{M}\tilde{\mathcal{M}} = \begin{bmatrix} |a_{11}|^2 - |a_{12}|^2 & -a_{11}a_{21}^* + a_{12}a_{22}^* \\ a_{21}a_{11}^* - a_{22}a_{12}^* & -|a_{21}|^2 + |a_{22}|^2 \end{bmatrix}, \tag{2.14}$$

$$\tilde{\mathcal{M}}\mathcal{M} = \begin{bmatrix} |a_{11}|^2 - |a_{21}|^2 & a_{11}^* a_{12} - a_{21}^* a_{22} \\ -a_{12}^* a_{11} + a_{22}^* a_{21} & -|a_{12}|^2 + |a_{22}|^2 \end{bmatrix}. \tag{2.15}$$

We see that to be hermitean both $\mathcal{M}\tilde{\mathcal{M}}$ and $\tilde{\mathcal{M}}\mathcal{M}$ must be diagonal. That is we must have

$$a_{11}a_{21}^* - a_{12}a_{22}^* = 0,$$
$$a_{11}^* a_{12} - a_{21}^* a_{22} = 0. \tag{2.16}$$

The two equations may be considered as linear equations on $a_{12}, a_{21}^*$. Since the corresponding determinant is $|a_{11}|^2 - |a_{22}|^2$, when $|a_{11}|^2 \neq |a_{22}|^2$ these equations have only a trivial solution $a_{12} = a_{21} = 0$. In such a case $a_{11}, a_{22}$ are arbitrary. For the trivial solution we obtain that $\mathcal{M}, \tilde{\mathcal{M}}$ are diagonal with arbitrary diagonal entries and $\mathcal{M} = \tilde{\mathcal{M}}$. Otherwise, $|a_{11}|^2 = |a_{22}|^2$ and

$$a_{21}^* = (a_{22}^*/a_{11})a_{12}. \tag{2.17}$$

Since $|a_{12}| = |a_{21}|$ then $\mathcal{M}\tilde{\mathcal{M}} = \tilde{\mathcal{M}}\mathcal{M}$ and they are proportional to the unit matrix. Therefore, all non-trivial $\mathcal{M}, \tilde{\mathcal{M}}$ are proportional to a $SU(1,1)$ matrix, provided . Hence all non-trivial positive mass solutions can be represented as

$$\mathcal{M} = mU_1 R U_2, \qquad U_{1,2} = diag(\exp i\alpha_{1,2}, \exp i\beta_{1,2}), \qquad \sum(\alpha_k + \beta_k) = 0,$$

$$R = \begin{pmatrix} c_\lambda & s_\lambda \\ s_\lambda & c_\lambda \end{pmatrix}, \quad c_\lambda = \cosh \lambda = a_{11}/\sqrt{|a_{11}|^2 - |a_{12}|^2}, \quad s_\lambda = \sinh \lambda, \tag{2.18}$$

where $m = \sqrt{|a_{11}|^2 - |a_{12}|^2}$. The matrix generates two equal real positive masses $m$ if $|a_{11}| > |a_{12}|$ and two equal zero masses if $|a_{11}| = |a_{12}|$. Otherwise the two masses are no longer real.

The mass degeneracy in the non-trivial case indicates the presence of what we, by analogy with isospin, call $SU(1,1)$ flavor symmetry and that the two flavors form a *flavor spin* doublet. To complete the classification we note that if $|a_{11}|^2 = |a_{12}|^2$ in the non-trivial case the two flavors are massless.

In summary, Lagrangian (2.5) with positive physical masses admits only specific mass matrices and describes either two particles with arbitrary masses, or one $SU(1,1)$ flavor doublet of mass $m$.

Let us now consider the four-dimensional case. In the $2 \times 2$ block notation

$$\mathcal{M} = \begin{bmatrix} A & B \\ C & D \end{bmatrix}, \qquad \tilde{\mathcal{M}} = \begin{bmatrix} A^+ & -C^+ \\ -B^+ & D^+ \end{bmatrix}.$$

We can use two unitary transformations from $U(2) \times U(2) \subset U(2,2)$ on the left and the right modes to diagonalize the diagonal $2 \times 2$ blocks $A, D$ in $\mathcal{M}, \tilde{\mathcal{M}}$ so that the diagonal entries are all positive. We can do this because the kinetic term in the action is invariant under $U(2) \times U(2)$. Then, returning to four-index notation, the conditions for a physical mass term amount to

$$a_{kk} a_{mk}^* - a_{km} a_{mm}^* = 0,$$
$$a_{kk}^* a_{kn} - a_{mk}^* a_{mm} = 0, \qquad (2.19)$$

which means either $|a_{kk}|^2 = |a_{mm}|^2$ or $a_{mk} = a_{kn} = 0$ for all elements in the upper right and lower left blocks. In the $2 \times 2$ block notation

$$\mathcal{M}\tilde{\mathcal{M}} = \begin{bmatrix} AA^+ - BB^+ & -AC^+ + BD^+ \\ CA^+ - DB^+ & -CC^+ + DD^+ \end{bmatrix},$$

$$\tilde{\mathcal{M}}\mathcal{M} = \begin{bmatrix} A^+A - C^+C & A^+B - C^+D \\ -B^+A + D^+C & -B^+B + D^+D \end{bmatrix}, \qquad (2.20)$$

where

$$A^+A = AA^+ = diag(a_1^2, a_2^2), \qquad DD^+ = D^+D = diag(d_1^2, d_2^2). \qquad (2.21)$$

The condition (2.11) implies the vanishing of the off-diagonal blocks

$$AC^+ - BD^+ = 0, \qquad A^+B - C^+D = 0. \qquad (2.22)$$

Since $A, D$ are diagonal with positive diagonal entries, the solution of (2.22) is

$$C = B^+. \qquad (2.23)$$

It then follows that $\mathcal{M}\tilde{\mathcal{M}} = \tilde{\mathcal{M}}\mathcal{M}$, which satisfies (2.12).

We see that in order to generate physical masses, the quasi-diagonalized mass matrix must be Hermitean. In the Standard Model $\tilde{\mathcal{M}} = \mathcal{M}^+$ and any matrix can be a mass matrix, because $\mathcal{M}\mathcal{M}^+$, $\mathcal{M}^+\mathcal{M}$ are hermitean and have equal positive eigenvalues. Not so here.

We can now classify all possible physical mass spectra in terms of elements of $A, B, D$. The mass squared eigenvalues are the eigenvalues of two matrices $AA^+ - BB^+$ and $DD^+ - B^+B$. Depending on how many entries in $AA^+, DD^+$ coincide a single bi-spinor field can generate from one to four different masses. The solutions can be organized in terms of the flavor symmetry of the Lagrangian: $SU(2,2)$ quadruplet when all masses coincide or $SU(1,1)$ for each flavor doublet when there are two pairs or only one pair of equal masses or none, when all masses are different. In the three-dimensional case we obtain a similar result, except that $SU(2,2)$ symmetry cannot be realized with only three masses. The mass spectrum in this case is composed of the spectrum of the two-dimensional case plus an arbitrary mass. For completeness we should mention the $SU(3,1)$ case. It cannot be realized in our formalism, because the assumption that fermionic degrees of freedom are described by quantum differential forms leads only to $SU(2,2)$ or $SU(2,1)$ cases.

### 3. Mass Eigenstate Basis and Flavor Mixing

Having classified the possible physical mass terms in the theory, we can describe the transition from the interaction to the mass eigenstate basis and the resulting flavor mixing. Let us begin by looking at the version of the SM, where some of $\mathcal{M}\mathcal{M}^+$ mass eigenvalues coincide. This would mean that all field components with the same eigenvalue transform in a unitary representation of $SU(3)$ or its subgroup and hence describe a single particle state. Depending on how many eigenvalues coincide, mass matrix would then describe from one to three particles. Note, that when some of its eigenvalues coincide, the diagonalization of $\mathcal{M}\mathcal{M}^+$ is no longer unique. It becomes unique only when all symmetries responsible for the equality of the eigenvalues are broken.

The same arguments apply in our case. For four generations Lagrangian with the kinematic $SU(2,2)$ symmetry the number of the particles the Lagrangian describes is determined by the symmetry of the mass matrix. Obviously, the same also applies to the kinematic $SU(2,1)$ or $SU(1,2)$ with three generations.

Let us begin with two generations. The starting point is the Lagrangian

$$\mathcal{L} = (\overline{\psi}^1, \overline{\psi}^2) \begin{pmatrix} i\partial\!\!\!/ - mc_\lambda & -ms_\lambda \gamma^5 \\ ms_\lambda \gamma^5 & -i\partial\!\!\!/ + mc_\lambda \end{pmatrix} \begin{pmatrix} \psi^1 \\ \psi^2 \end{pmatrix}. \tag{3.1}$$

Like in the degenerate mass SM we need to transform it into the form where all mass eigenvalues become explicit and equations of motion decouple. This can be achieved with the transformation to what we call the *chiral flavor spin* variables

$$\begin{pmatrix} \psi^1 \\ \psi^2 \end{pmatrix} = T \begin{pmatrix} \xi^1 \\ \xi^2 \end{pmatrix}, \qquad T = \frac{1}{\sqrt{2}} \begin{pmatrix} 1 & -1 \\ 1 & 1 \end{pmatrix}. \tag{3.2}$$

We use the term because for two dimensions in the Dirac representation $\gamma^0 = \Gamma_2$ and (3.2) is the transformation to the representation where $\gamma^3$ is diagonal. The same applies to four dimensions. There $\gamma^0 = \Gamma$ and, after 4D version of (3.2), it is $\gamma^5$ becomes diagonal. Applying (3.2) to (3.1)

we obtain

$$\mathcal{L} = (\bar{\xi}^1, \bar{\xi}^2) \begin{pmatrix} 0 & (i\partial - mc_\lambda - ms_\lambda \gamma^5) \\ (i\partial - mc_\lambda + ms_\lambda \gamma^5) & 0 \end{pmatrix} \begin{pmatrix} \xi^1 \\ \xi^2 \end{pmatrix}, \quad (3.3)$$

leading to the decoupled equations of motion for each component. The equations become identical for $c_\lambda = 1, s_\lambda = 0$. In the *chiral flavor spin* basis we can replace $\gamma^5$ with $\pm 1$ to obtain

$$i\partial \xi^1{}_{R,L} - m e^{\pm \lambda} \xi^1{}_{L,R} = 0,$$

$$i\partial \xi^2{}_{R,L} - m e^{\mp \lambda} \xi^2{}_{L,R} = 0 \quad (3.4)$$

We see that each field component $\xi^k$ satisfies the Klein-Gordon equation $\left((i\partial)^2 - (m)^2\right)\xi^k = 0$. This follows from the identity $(i\partial - m(c_\lambda \pm s_\lambda \gamma^5))(i\partial - m(-c_\lambda \pm s_\lambda \gamma^5)) = (i\partial)^2 - (m)^2$. Therefore, each doublet component formally has the equations of motion of a free particle. Note that the *chiral flavor spin* transformation does not break the $SU(1,1)$ symmetry of (3.1). Instead, it replaces it by an isomorphic symmetry. Despite the unorthodox appearance, the Lagrangian (3.3) results in the Hamiltonian that is diagonal in the creation-annihilation (*c-a*) operators, provided we use a compensating unitary transformation with $T^+$ applied to the *c-a* operators. We can use $T^+$, because for anticommuting *c-a* operators unitary rotations preserve the anti-commutation relations.

Just like in the degenerate mass version of the SM, to obtain from (3.3) a physical particle state for each field component, the $SU(1,1)$-related symmetry must be broken. In fact, it must broken by quantum effects, because $SU(1,1)$ is a subgroup of $SU(2,2)$, which is isomorphic to the conformal group $C(1,3) \approx SO(2,4)$. The explicit construction of the isomorphism can be found in [13]. It is well-known that the conformal group of the Minkowski space-time is broken via the trace anomaly.

In our case the breaking the $SU(1,1)$ symmetry fixes the $SU(1,1)$ gauge uniquely, because we can obtain a renormalized theory with physical mass states only if $c_\lambda = 1, s_\lambda = 0$. This follows from the results of the preceding section, where we have shown that the mass term must either be proportional to either a $SU(1,1)$ element or the unit matrix. From the form of (3.3) we observe that the one-loop renormalization corrections to the mass term must be diagonal

$$\Delta \mathcal{L}_m = (\bar{\xi}^1, \bar{\xi}^2) \begin{pmatrix} \Delta m & 0 \\ 0 & \Delta m \end{pmatrix} \begin{pmatrix} \xi^1 \\ \xi^2 \end{pmatrix}. \quad (3.5)$$

This results in two different masses $m_{1,2} = \Delta m \pm m$. These leading contributions come from the one-loop self-energy diagrams for $\xi^1, \xi^2$. Off-diagonal contributions appear first from two-loop diagrams.

The $SU(1,1)$ case is generalized to four generations $SU(2,2)$ or three generations $SU(2,1)$ in a straightforward way. In summary, for four generations despite non-compactness of $SU(2,2)$ the transition from the interaction to the mass basis is effected by a unitary transformation using block diagonal $U(2) \times U(2)$. The same happens for $SU(2,1)$ using $U(2) \times U(1)$. Let us consider these cases in more detail.

It follows from the above discussion that in the $SU(2,2)$ case the transformation to the mass basis involves a block-diagonal unitary transformation that mixes $\xi^1, \xi^2$ and separately $\xi^3, \xi^4$

$$U = \begin{bmatrix} W_1 & 0 \\ 0 & W_2 \end{bmatrix}, \quad W_k \in U(2), \tag{3.6}$$

possibly followed by the unitary transformations $T$ in (3.2) that mix the members of the $k$-th $SU(1,1)$ pairs, such as $\xi^1, \xi^3$ and/or $\xi^2, \xi^4$.

## 4. The CKM Mixing Matrix

We now consider flavor mixing in our theory, beginning with the quark sector. For the SM Higgs field expectation value $\phi = (0, v/\sqrt{2})$ and the left mass terms diagonalized as $M^{u,d} = (v/\sqrt{2}) V_L^{u,d+} M_{diag}^{u,d} V_R^{u,d}$ the CKM matrix $V_{CKM}$ is defined by $V_{CKM} = V_L^u V_L^{d+}$ and similarly for the PMNS matrix $U_{PMNS}$. Combining (3.2) and (3.6) we obtain all possible charged current quark mixing matrices for the four generations SM

$$V^4 = (T_{kl} U_U)(T_{mn} U_D)^+, \tag{4.1}$$

where $T_{kl}, k, l = 0,1$ are obtained by combining either the $2 \times 2$ unit matrix ($k = 0$ or $l = 0$) or a transformations (3.2) ($k = 1$ or $l = 1$) acting on pairs of indices $i, j = 1,3$ and $i, j = 2,4$ of the $4 \times 4$ matrix. There are four possible $T_{kl}, k, l = 1,2$ and hence 16 possible quite distinct textures of $V$ parameterized by two $U(2)$ matrices $\hat{W}_1 = W_1 W_1'$ and $\hat{W}_2 = W_2 W_2'$ in (3.6). The choice from which one obtains the closest match with the experimental values for the three generation $V_{CKM}$ is

$$V_{CKM}^4 = T_{11} U T_{11}^+, \quad U = U_U U_D^+. \tag{4.2}$$

Explicitly

$$V_{CKM}^4 = \begin{bmatrix} 1/\sqrt{2} & 0 & -1/\sqrt{2} & 0 \\ 0 & 1/\sqrt{2} & 0 & -1/\sqrt{2} \\ 1/\sqrt{2} & 0 & 1/\sqrt{2} & 0 \\ 0 & 1/\sqrt{2} & 0 & 1/\sqrt{2} \end{bmatrix} \begin{bmatrix} x_1 & y_1 & 0 & 0 \\ z_1 & w_1 & 0 & 0 \\ 0 & 0 & x_2 & y_2 \\ 0 & 0 & z_2 & w_2 \end{bmatrix} \begin{bmatrix} 1/\sqrt{2} & 0 & 1/\sqrt{2} & 0 \\ 0 & 1/\sqrt{2} & 0 & 1/\sqrt{2} \\ -1/\sqrt{2} & 0 & 1/\sqrt{2} & 0 \\ 0 & -1/\sqrt{2} & 0 & 1/\sqrt{2} \end{bmatrix}, \tag{4.3}$$

where $x_k$, *etc* are the matrix values in (3.6). The form of the $V_{CKM}^4$ indicates that before the $SU(1,1)$ symmetry breaking the four generations of up and down quarks form two $SU(1,1)$ doublets. The final form of $V_{CKM}^4$ is

$$V_{CKM}^4 = \begin{bmatrix} x_+ & y_+ & x_- & y_- \\ z_+ & w_+ & z_- & w_- \\ x_- & y_- & x_+ & y_+ \\ z_- & w_- & z_+ & w_+ \end{bmatrix}, \tag{4.4}$$

where $x_\pm = (x_1 \pm x_2)/2$ etc.

To obtain the 3 generation mixing matrix we need to assume $W_1 \approx W_2$ and either (1) non-observation of one of the four generations because of its high mass or (2) decoupling of one of the generations from all SM gauge fields or (3) near-degeneracy in mass of the two heaviest generations. We then obtain the three-generation CKM matrix derived in [5]

$$V_{CKM} = \begin{bmatrix} x_+ & y_+ & y_- \\ z_+ & w_+ & w_- \\ z_- & w_- & w_+ \end{bmatrix}. \tag{4.5}$$

This matrix has two real parameters and one rephasing non-removable phase, which is one real parameter less then in the SM. Note that inequality of $a_{31} = z_- = (z_1 - z_2)/2$, $a_{13} = y_- = (y_1 - y_2)/2$ is due to the possible difference in the relative phases $(\eta_1 - \eta_2)$ of $y_1 = \sin\theta_1 e^{i\eta_1}$ relative to $y_2 = \sin\theta_2 e^{i\eta_2}$ and $(\xi_1 - \xi_2)$ of $z_1 = \sin\theta_1 e^{i\xi_1}$ relative $z_2 = \sin\theta_2 e^{i\xi_2}$. These relative phases cannot be eliminated by re-phasing of the fermionic fields.

From (4.5), taking into account that the phases of the matrix elements can be changed by re-phasing the fields, we obtain two tree level predictions of equality of two pairs of its elements:

$$|V_{cb}| = |V_{ts}|, \tag{4.6}$$

$$|V_{cs}| = |V_{tb}|. \tag{4.7}$$

$|V_{cb}| = |V_{ts}|$ is experimentally observed to lie within 1 $\sigma$ errors for $|V_{cb}|, |V_{ts}|$ [1]. $|V_{cb}| = (40.8 \pm 1.4) \times 10^{-3}$ $|V_{ts}| = (41.5 \pm 0.9) \times 10^{-3}$. $|V_{cs}| = |V_{tb}|$ is also observed but with less accuracy. The values are measured at $|V_{cs}| = 0.975 \pm 0.006$ and $|V_{tb}| = 1.014 \pm 0.029$ so their 2 $\sigma$ ranges overlap.

We will now discuss three scenarios of the non-observation of the fourth quark generation that we mentioned above. The first and most widely discussed one is that the masses of $t', b'$ are too large for them to be generated experimentally. Currently the lower limits are $m_{t'} > 1.3 \times 10^3 GeV, m_{b'} > 1.6 \times 10^3 GeV$ [1]. The second one is that the fourth generation decouples from all gauge fields and interacts only gravitationally, thus acting as dark matter. The third one apparently has not been mentioned in the literature so far. Namely, that the fourth generation is not observed, because of near degeneracy in mass with the third generation. In the SM there is no reason for such degeneracy. However, in our theory our first and second generations and separately the second and the fourth generation form $SU(1,1)$ doublets before its breakdown. Because of this, is would not be extraordinary to find that the mass differences between the third and the fourth generations should be similar to that of the first and the second generations. If masses of $t',b'$ lie within $1\sigma$ error ranges of the measured $t,b$ masses, they would not be distinguishable experimentally from $t',b'$.

The experimentally measured masses of $t,b$ quarks are $m_b = 4.8^{+0.03}_{-0.02} GeV$ and depending on its definition ranges from $m_t = 162.5^{+2.1}_{-1.5} GeV$ to $m_t = 173.2 \pm 0.9 GeV$. The corresponding mass differences are $m_c - m_u \approx 1.7 GeV$, $m_s - m_d \approx 93 MeV$. It is not so implausible that $|m_t - m_{t'}|$ and $|m_b - m_{b'}|$ lie in the range of $1 GeV$ and $30 MeV$ respectively. Therefore, this scenario needs to be taken as a possible explanation.

Let us consider the effects of the three different scenarios on the mixing matrix. The effects of the first two are identical but differ from those in the third scenario. The difference lies in the statistics of the event count. Whereas in the first two scenarios the fourth generation is not present in the event count, in the third scenario it is but is summed together with the third generation event count.

As a result the first two scenarios result in both horizontal and the vertical unitarity deficits. The first row and the first column possible deficits were discussed in [14]. The third scenario produces a different effect. First, in this scenario the definition of the effective three generation matrix would depend on the choice of the incoming states. For simplicity we will assume that all incoming states are the up quark states and that the third and the fourth generations of the up quark states are produced at the same rate. Note that we have to take into account that as quantum states $t',b'$ are distinguishable from $t,b$, respectively. Hence, the rule that we have to average the amplitudes over the incoming states and sum them over the outgoing states now applies to the probabilities and not to the amplitudes. If we denote as $\hat{t},\hat{b}$ the effective top/bottom quarks in the effective three-generation mixing matrix, then summing over the final states we obtain

$$|V_{u\hat{b}}|^2 = |V_{ub}|^2 + |V_{ub'}|^2 \tag{4.8}$$

$$|V_{c\hat{b}}|^2 = |V_{cb}|^2 + |V_{cb'}|^2 \tag{4.9}$$

$$|V_{t\hat{b}}|^2 = |V_{tb}|^2 + |V_{tb'}|^2 \tag{4.10}$$

$$|V_{t'\hat{b}}|^2 = |V_{t'b}|^2 + |V_{t'b'}|^2 \tag{4.11}$$

At the same time to obtain $|V_{\hat{t}\hat{b}}|^2$ we have to average over the incoming $t, t'$, which produces

$$|V_{\hat{t}\hat{b}}|^2 = \frac{1}{2}\left(|V_{tb}|^2 + |V_{tb'}|^2 + |V_{t'b}|^2 + |V_{t'b'}|^2\right). \tag{4.12}$$

We see that because of the averaging the vertical unitarity can be violated. For example, using the four-generation unitarity we obtain

$$|V_{u\hat{b}}|^2 + |V_{c\hat{b}}|^2 + |V_{\hat{t}\hat{b}}|^2 = 1 + \frac{1}{2}\left(|V_{ub}|^2 + |V_{cb}|^2 + |V_{ub'}|^2 + |V_{cb'}|^2\right) = 1 + \Delta_3 \geq 1, \tag{4.13}$$

which exhibits a small positive unitarity deficit. The remaining two vertical unitarity sums are given by

$$|V_{ud}|^2 + |V_{cd}|^2 + |V_{\hat{t}d}|^2 = 1 - \frac{1}{2}\left(|V_{td}|^2 + |V_{t'd}|^2\right) = 1 + \Delta_1 \leq 1, \tag{4.14}$$

$$|V_{us}|^2 + |V_{cs}|^2 + |V_{\hat{t}s}|^2 = 1 - \frac{1}{2}\left(|V_{ts}|^2 + |V_{t's}|^2\right) = 1 + \Delta_2 \leq 1. \tag{4.15}$$

Here we have small negative unitarity deficits. At the same time the horizontal unitarity obviously holds by construction. Using four-dimensional unitarity, we can rewrite

$$\Delta_3 = 1 - \frac{1}{2}\left(|V_{td}|^2 + |V_{t'd}|^2 + |V_{ts}|^2 + |V_{t's}|^2\right),$$

and summing all of the deficits we obtain

$$\sum_k \Delta_k = 0. \tag{4.16}$$

The positive and the negative deficits cancel each other because of the horizontal unitarity. This relation between the unitarity deficits in principle can be used to determine experimentally whether the first two or the third scenarios are realized in the experiments.

Another important fact about the unitarity deficits is that their magnitude is of the order of the largest small mixing element. It is worth noting that the possible unitarity deficits reported in [14] are in fact of that order.

## 5. The PMNS Mixing Matrix

Let us turn to the leptonic flavor mixing. In the case of leptons the assumption of the non-interacting fourth lepton generation can be used to derive the three generation mixing like for quarks but it turns out to be not necessary. Here it is enough to assume that there is only one broken $SU(1,1)$ doublet consisting of *mu* and *tau* leptons. The rest of the leptons are $SU(1,1)$ singlets. This implies that the classical Lagrangian has $SU(1,1)$ *mu - tau* symmetry. It is worth noting that a possible *mu - tau* symmetry and its consequences have been discussed in the literature [15].

Using the mass matrix diagonalizing unitary transformations described above we obtain for leptons

$$U_{PMNS} = T_E U_E U_N^+ T_N^+, \tag{5.1}$$

An essentially unique choice for $U_{PMNS}$ that fits the experimental data is

$$T_E = \begin{bmatrix} 1 & 0 & 0 \\ 0 & 1/\sqrt{2} & -1/\sqrt{2} \\ 0 & 1/\sqrt{2} & 1/\sqrt{2} \end{bmatrix}, U_E = \begin{bmatrix} x_1 & 0 & y_1 \\ 0 & u_1 & 0 \\ z_1 & 0 & w_1 \end{bmatrix}, U_N = \begin{bmatrix} x_2 & y_2 & 0 \\ z_2 & w_2 & 0 \\ 0 & 0 & u_2 \end{bmatrix}, T_N = 1, \tag{5.2}$$

where $x_k, \ldots, w_k$ are elements of $U(2)$ and $u_k$ are complex phases. All other choices either do not fit the experimental data or are obtained by renaming the lepton generations and assigning different signatures to the *mu-tau* doublet. Explicitely we have

$$U_{PMNS} = \begin{bmatrix} x_1 x_2 & y_1 u_2 & x_1 y_2 \\ (z_1 x_2 + u_1 z_2)/\sqrt{2} & (z_1 y_2 + u_1 w_2)/\sqrt{2} & w_1 u_2/\sqrt{2} \\ (z_1 x_2 - u_1 z_2)/\sqrt{2} & (z_1 y_2 - u_1 w_2)/\sqrt{2} & w_1 u_2/\sqrt{2} \end{bmatrix}. \tag{5.3}$$

We observe that absolute values of two elements of the matrix are equal, namely

$$|U_{\mu 3}| = |U_{\tau 3}|, \tag{5.4}$$

and that it reduces to the tri-bimaximal (TBM) mixing matrix when $U_E = 1$. If we now set all $x_k, \ldots, w_k$, $u_1, u_2$ to be real positive except for $y_1 = -z_1^* = \sin\theta_{13} e^{-i\delta}$ then, after phases adjustment due to change of $\theta_{23} \to -\theta_{23}$, in (5.2) we recognize the standard parameterization of $U_{PMNS}$. Therefore, our theory predicts that at tree level

$$\sin\theta_{23} = \cos\theta_{23}. \tag{5.5}$$

The remaining two angles $\theta_{12}$, $\theta_{13}$ remain arbitrary.

It is worth to compare (5.3) with the matrix that is obtained by the reduction from four to three generations as was done above for quarks. The essentially unique fit with the lepton mixing data texture comes from using

$$U_{PMNS}^{4} = \begin{bmatrix} 1 & 0 & 0 & 0 \\ 0 & 1/\sqrt{2} & 0 & -1/\sqrt{2} \\ 0 & 0 & 1 & 0 \\ 0 & 1/\sqrt{2} & 0 & 1/\sqrt{2} \end{bmatrix} \begin{bmatrix} x_1 & y_1 & 0 & 0 \\ z_1 & w_1 & 0 & 0 \\ 0 & 0 & x_2 & y_2 \\ 0 & 0 & z_2 & w_2 \end{bmatrix} \begin{bmatrix} \hat{x}_1 & 0 & \hat{y}_1 & 0 \\ 0 & \hat{x}_2 & 0 & \hat{y}_2 \\ \hat{z}_1 & 0 & \hat{w}_1 & 0 \\ 0 & \hat{z}_2 & 0 & \hat{w}_2 \end{bmatrix}, \tag{5.6}$$

which, after decoupling of the fourth generation, reduces to

$$U_{PMNS}^{3} = \begin{bmatrix} x_1\hat{x}_1 & y_1\hat{x}_2 & y_1\hat{y}_2 \\ (z_1\hat{x}_1 - z_2\hat{z}_1)/\sqrt{2} & (w_1\hat{x}_2 - w_2\hat{z}_2)/\sqrt{2} & (w_1\hat{y}_2 - w_2\hat{w}_2)/\sqrt{2} \\ (z_1\hat{x}_1 + z_2\hat{z}_1)/\sqrt{2} & (w_1\hat{x}_2 + w_2\hat{z}_2)/\sqrt{2} & (w_1\hat{y}_2 + w_2\hat{w}_2)/\sqrt{2} \end{bmatrix}. \tag{5.7}$$

Also this mixing matrix exhibits constraints on its elements. However they are no longer linear.

From [1] we find the present $3\sigma$ range for the leptonic angle $\theta_{23}$ derived from all available up to 2022 data and for various neutrino mass orderings is $(0.41 - 0.61)$. The best fit plus $1\sigma$ error values for $\sin^2\theta_{23}$ vary from $0.451^{+0.019}_{-0.016}$ to $0.578^{+0.016}_{-0.021}$, depending on inclusion or exclusion of the SK atmospheric data.

As we have shown in our theory the difference between quark and lepton mixing comes primarily from $SU(1,1)$ *flavor spin* doublet assignments. There are four such doublets for quarks and only one for leptons. This may not answer the question why such a difference arises, but allows looking at the problem from a somewhat different angle. Another significant difference with the quark case is that, whereas for quarks the sign of the kinetic term is the same for the members of the isospin doublet, for leptons for the second and the third generation of leptons it is the opposite, so that while for *mu* and *tau* the signs are plus and minus, respectively, for the *mu* and *tau* neutrinos it is minus and plus. It would have been a potential problem for our theory, were not both *mu-tau* flavor spin and isospin symmetries broken, as is the case.

## 6. Quantization of the Anti-Dirac Spinors

In the preceding section we showed that using the spinors whose action is the negative of the standard ones helps to explain mixing matrix textures and near coincidence of their elements both for quarks and leptons. The free field action for such spinors is given by

$$S_{aD} = -\int d^4x\, \overline{\psi}\, (i\partial\!\!\!/ - m)\psi. \tag{6.1}$$

We shall call such fields the *anti-Dirac* or *aD*-spinor fields. aD spinors have not been used in the model building, because their standard quantization obviously results in the Hamiltonian unbound from below. In this section we describe the modification of quantization of such fields that solves this problem.

Although the equations of motion for both D and aD fields are the same, the minus in (6.1) causes non-trivial physical consequences. The most prominent of them is the change of sign of the classical Hamiltonian: $H_{aD} = -H_D$. Recall that it is only after the standard quantization that the operator $H_D$ becomes bounded from below. Hence, the standard quantization of $H_{aD}$ results in an unphysical theory.

To understand how to cure the problem, recall the expression for the evolution for operator $A(t)$ from its initial state $A(t_i)$ to the final state $A(t_f)$ in the Heisenberg representation

$$A(t_f) = \exp(iH(t_f - t_i))A(t_i)\exp-(iH(t_f - t_i)). \tag{6.2}$$

If we replace $H$ in (6.2) with $-H$, we see that the evolution of $A(t)$ is completely equivalent to evolution with the original $H$ but backwards in time. Obviously, the same applies to the interaction representation and hence to the expression for the S-matrix. We conclude that S-matrix with a negative definite Hamiltonian is equivalent to the S-matrix with the time-reversed evolution. We can use this equivalence to define a well-defined quantum field theory, where some of the particles are described by (6.1).

To proceed further it is helpful to use quantum filed theory formalism, where time evolution in both directions enters on equal footing. Such extended formalism has been described by Keldysh [16]. Although it is mostly used in the solid state physics to describe non-equilibrium QFT at finite temperature, it can very well be used at zero temperature if we omit the averaging over thermal states and set the initial density matrix to one.

In the Keldysh QFT formalism one considers the evolution of Greens function as a T-product along the time path that extends from $-\infty$ to $+\infty$ and then back to $-\infty$. As a result one can construct Greens functions that are the vacuum expectation values of the fields with time coordinates on both positive and the negative time branches. This is a consistent procedure if we consider the time coordinates on the negative time branch as lying in the absolute future of all time values on the positive branch.

As a result the definition of various two point Greens functions and their perturbation theory can be considered from a unified point of view. The matrix of four Greens functions is assembled into a single one

$$G(x,y) = \begin{bmatrix} G^c(x,y) & G^-(x,y) \\ G^+(x,y) & \tilde{G}^c(x,y) \end{bmatrix}, \tag{6.3}$$

$$G^c(x,y) = -i\langle 0|T\psi(x_+)\overline{\psi}(y_+)|0\rangle, \qquad x_\pm = (t_\pm, \vec{x}),$$

$$\tilde{G}^c(x,y) = -i\langle 0|\tilde{T}\psi(x_-)\overline{\psi}(y_-)|0\rangle, \qquad (6.4)$$

$$G^\pm(x,y) = -i\langle 0|T\psi(x_\pm)\overline{\psi}(y_\mp)|0\rangle,$$

where $t_+$ is the point on the positive time direction path branch while $t_-$ is a point on the negative time direction path branch. $\tilde{T}$ is the T-product on the negative time path branch, that is with respect to negative time direction. All coupling constants acquire a minus sign on the negative time branch and symbolically can be assembled into

$$g \to \begin{bmatrix} g & 0 \\ 0 & -g \end{bmatrix}. \qquad (6.5)$$

Using these generalized Greens functions, not all of them independent, we can construct a well-defined interaction representation, where $G(x,y)$ enters as the main object.

We observe that the aD quantum field described by (6.1) for all intends and purposes is the quantum field of the Keldysh formalism that is defined on the negative time direction branch. We can use this observation to define a theory, where aD fields are incorporated in the standard QFT with the associated S-matrix. To do this we need to make sure that all scattering processes are employing exchange of particles with positive energy.

We now describe how to achieve this. Recall the standard mode expansion for the Dirac field

$$\psi(x) = \int \frac{d^3k}{(2\pi)^3} \frac{m}{k^0} \left( b_r(\vec{k}) u^r(\vec{k}) e^{-ikx} + d_r^+(\vec{k}) v^r(\vec{k}) e^{ikx} \right) \qquad (6.6)$$

$$\overline{\psi}(x) = \int \frac{d^3k}{(2\pi)^3} \frac{m}{k^0} \left( b_r^+(\vec{k}) \overline{u}^r(\vec{k}) e^{ikx} + d_r(\vec{k}) \overline{v}^r(\vec{k}) e^{-ikx} \right), \qquad (6.7)$$

where $k^0 = +\sqrt{\vec{k}^2 + m^2}$. The plane-wave solutions $u^r(\vec{k})$, $r = 1,2$, for the positive and $v^r(\vec{k})$, $r = 1,2$, for the negative energy satisfy $(\slashed{k} - m)u^r(\vec{k}) = (\slashed{k} + m)v^r(\vec{k}) = 0$ and are normalized in the standard way: $\overline{u}^p(\vec{k})u^q(\vec{k}) = -\overline{v}^p(\vec{k})v^q(\vec{k}) = \delta^{pq}$. After quantization $b_r^+(\vec{k})$, $b_s(\vec{k})$ become the anti-commuting c-a operators for the Dirac particles, while $d_r^+(\vec{k})$, $d_s(\vec{k})$ become the c-a operators for the Dirac antiparticles. The energy momentum operator $P^\mu$ is

$$P^\mu =: \int \frac{d^3k}{(2\pi)^3} \frac{m}{k^0} k^\mu \left( b_r^+(\vec{k}) b_r(\vec{k}) + d_r^+(\vec{k}) d_r(\vec{k}) \right):, \qquad \langle 0|P^\mu|0\rangle \geq 0, \qquad (6.8)$$

where : : denotes the normal ordering of the operators.

To define S-matrix one uses the time-ordered product of two spinor fields,

$$T\psi(x)\bar{\psi}(y) = \theta(x^0 - y^0)\psi(x)\bar{\psi}(y) - \theta(y^0 - x^0)\bar{\psi}(y)\psi(x), \tag{6.9}$$

and the Feynman propagator for the Dirac field

$$S_F(x-y) = -i\langle 0|T\psi(x)\bar{\psi}(y)|0\rangle = \int \frac{d^4k}{(2\pi)^4} \frac{(\slashed{k}+m)}{k^2 - m^2 + i\varepsilon} e^{-ik(x-y)}. \tag{6.10}$$

For the time-reversed propagator with $\tilde{T}$ we obtain

$$\tilde{S}_F(x-y) = -i\langle 0|\tilde{T}\psi(x)\bar{\psi}(y)|0\rangle = \int \frac{d^4k}{(2\pi)^4} \frac{(\slashed{k}+m)}{k^2 - m^2 - i\varepsilon} e^{-ik(x-y)}, \tag{6.11}$$

where we took into account that the contour of integration for $\tilde{S}_F(x-y)$ is in the opposite direction to that of $S_F(x-y)$, which generates a factor of $-1$ in (6.11). Note the change of path around the pole due to replacement of $+i\varepsilon$ for $S_F(x-y)$ to $-i\varepsilon$ for $\tilde{S}_F(x-y)$. It is indicative of the revered time propagation or more precisely it indicates propagation of positive energy into the past and consequently the negative energy into the future.

Clearly we cannot use the mode expansion (6.6-7) for the aD spinor field, because then we obtain a negative definite Hamiltonian operator. Furthermore, since the aD field in effect is defined on the negative time direction branch, for its description we need to use the reversed time order T-product definition. This is a physically understandable result: $\tilde{S}_F(x)$ is the amplitude of the process, where a particle with positive energy first destroyed and then created. This is equivalent to saying that a particle with negative energy as first created and then destroyed. In both processes the flow of positive energy is backwards in time as is expected on the negative direction time branch.

The resolution of the negative Hamiltonian problem comes if we use the fact that in the mode expansion (6.6-7) it is left to our choice what we declare as particles and what as anti-particles. Therefore, if we exchange the creation and the annihilation operators in (6.6-7), then the corresponding amplitude will describe positive energy transfer in the positive time direction and we obtain $\langle 0|P_{aD}{}^\mu|0\rangle \geq 0$. In summary, the fields in (6.1) with the swapped c-a operators propagate backwards in time apparently but they transfer positive energy in the positive time direction. It may be useful here to distinguish between the kinematic direction of time and the causal direction of time. If the kinematic direction of time is defined by the sign of the action, the causal direction of time is defined by the direction of the positive energy transfer. Then kinematically aD fields propagate in the negative time direction, while causally the time direction of their evolution is positive.

With all the above in mind, the modified mode expansion that ensures the positivity of the Hamiltonian and maintains the causality becomes

$$\psi_{aD}(x) = \int \frac{d^3k}{(2\pi)^3} \frac{m}{k^0} \left( b_r^+(\vec{k}) u^r(\vec{k}) e^{-ikx} + d_r(\vec{k}) v^r(\vec{k}) e^{ikx} \right), \quad (6.12)$$

$$\overline{\psi}_{aD}(x) = \int \frac{d^3k}{(2\pi)^3} \frac{m}{k^0} \left( b_r(\vec{k}) \overline{u}^r(\vec{k}) e^{ikx} + d_r^+(\vec{k}) \overline{v}^r(\vec{k}) e^{-ikx} \right), \quad (6.13)$$

where now $b_r^+(\vec{k})$, $b_s(\vec{k})$ become the c-a operators for aD anti-particles, while $d_r^+(\vec{k})$, $d_s(\vec{k})$ are the c-a operators for aD particles.

To computing the vacuum expectation value of the T-product of the aD fields with mode expansion (6.12-13), we define

$$S_{aF}(x) = -i\tilde{T}\psi_{aD}(x)\overline{\psi}_{aD}(y), \quad (6.14)$$

$$\tilde{T}\psi_{aD}(x)\overline{\psi}_{aD}(y) = \theta(y^0 - x^0)\psi_{aD}(x)\overline{\psi}_{aD}(y) - \theta(x^0 - y^0)\overline{\psi}_{aD}(y)\psi(x). \quad (6.15)$$

After a simple calculation we obtain

$$S_{aF}(x) = \int \frac{d^4k}{(2\pi)^4} \frac{(\slashed{k} + m)}{k^2 - m^2 + i\varepsilon} e^{-ikx}, \quad (6.16)$$

or not surprisingly

$$S_{aF}(x) = S_F(x). \quad (6.17)$$

If we now turn to the definition of the S-matrix we find that the same assignment of the c-a operators that changed the sign of the Hamiltonian works the same way on the interaction Hamiltonian. It in effect changes the sign of the coupling constant in the interaction Hamiltonian on the reversed time branch of the Keldysh path. If we look at the definition of S-matrix for fermions interacting with gauge fields given by

$$S = T \exp\left[ i\int_{-\infty}^{+\infty} d^4x \, \mathcal{L}_I(\psi(x), A(x)) \right], \quad (6.18)$$

where $\mathcal{L}_I(x)$ is the interaction Lagrangian, for generically $\mathcal{L}_I(\psi(x), A(x)) = g\overline{\psi}(x)A(x)\psi(x)$ we obtain that because $\mathcal{L}_I(x)$ is bilinear in spinor fields, swapping the c-a operators in their mode expansion after normal ordering results in effect in $g \to -g$ on the reversed time branch. This makes both coupling constants in (6.5) to have the same sign. Naturally, our arguments have to be confirmed in the renormalized theory. This, however, is out of scope for this article.

In summary, when we choose the appropriate mode expansion for the aD fields, the corresponding particles, just like the Dirac field particles, carry the positive energy forwards in time, despite the formally negative Hamiltonian derived from (6.1) and the negative gauge field coupling constant. This means that all amplitudes generated with aD fields are indistinguishable from those coming from the Dirac fields.

## 7. Summary

In summary, we have described here the consequences of a single assumption that the quantum fermionic degrees of freedom in the three or four generation theories are described by inhomogeneous differential forms. The assumption results in the change of the signature of the kinematic quadratic form to a non-Euclidean.

The most important consequence of the assumption is the change of the signature of the kinematic Lagrangian in such theories. The $SU(3)$ or $SU(4)$ invariant signature for three or four generations is replaced by $SU(2,1)$ or $SU(2,2)$ invariant signatures. The consequence of this is that for such Lagrangians not all Yukawa coupling terms generate physical real positive masses.

To extract the physical eigenstates one has to perform two $SU(2)$ transformations in the flavor space. Some of them are fixed $\pi/2$ rotations. Because of this the resulting mixing matrix has one real parameter less then in the SM. As a result, four elements of $V_{CKM}$ must be pairwise equal at the tree level. For $U_{PMNS}$ derived under the three generation assumption we obtain the tree level prediction that $\sin\theta_{23} = \cos\theta_{23}$.

We compared the quark and lepton mixing matrices we obtained with the latest experimental data [1] and found that in the quark sector $|V_{cb}| = (40.8 \pm 1.4) \times 10^{-3}$, $|V_{ts}| = (41.5 \pm 0.9) \times 10^{-3}$, while $|V_{cs}| = 0.975 \pm 0.006$ and $|V_{tb}| = 1.014 \pm 0.029$. In the lepton sector we have $\theta_{23}$ the 3 $\sigma$ range is $(0.41 - 0.61)$ and the best fit plus 1 $\sigma$ error values vary from $0.451^{+0.019}_{-0.016}$ to $0.578^{+0.016}_{-0.021}$.

To obtain three generation quark mixing we had to assume that in fact at least four generations of quark are present in nature. We discussed three scenarios for non-observation of the fourth generation. In the addition of the conventional high mass scenario, we considered the decoupling of the fourth generation or the mass near degeneracy of the third and the fourth generations. We showed that the mass near degeneracy scenario is observationally distinct from the first two. The lepton sector mixing can be derived either with three or with four generation assumptions. In the first case the tree level equality $\sin\theta_{23} = \cos\theta_{23}$ is predicted. In both sectors one obtains experimentally observable mixing matrix textures.

Finally, to be able to construct physically acceptable gauge theory for aD spinors, whose action is the negative of that for the Dirac spinors, we described a modification of the standard quantum field theory to include fermionic fields that propagate backwards in time kinematically but forwards in time causally. The modification implies no changes in the S-matrix for such particles compared to the standard theory of gauge interactions with gauge fields coupling to the fermion bilinears.